\documentclass[namedreferences]{solarphysics}

\usepackage[hyperref,optionalrh,solaromanenum]{spr-sola-addons} 
\usepackage{graphicx}        
\usepackage{color}           
\usepackage{breakurl}        

\newenvironment{disc}[1][Disclosure of Potential Conflicts of Interest]{\footnotesize\paragraph*{#1}}{}

\usepackage{spr-sola-capfix}

\chardef\us=`\_

\begin{document}

\begin{article}

\begin{opening}

\title{Coronal Holes and Open Magnetic Flux over Cycles 23 and 24}

%
\author[addressref={aff1,aff2},corref,email={chris.lowder@durham.ac.uk}]{\inits{C}\fnm{Chris}~\lnm{Lowder}\orcid{0000-0001-8318-8229}}
\author[addressref={aff2},corref]{\inits{}\fnm{Jiong}~\lnm{Qiu}}
\author[addressref={aff2,aff3},corref]{\inits{}\fnm{Robert}~\lnm{Leamon}\orcid{0000-0002-6811-5862}}

%
\runningauthor{Lowder \textit{et al.}}
\runningtitle{Coronal Holes in Cycles 23 and 24}

\address[id=aff1]{Department of Mathematical Sciences, Durham University, Durham, DH1 3LE, UK}
\address[id=aff2]{Department of Physics, Montana State University, Bozeman, MT 59717, USA}
\address[id=aff3]{Department of Astronomy, University of Maryland, College Park, MD 20742, USA}

\begin{abstract}
As the observational signature of the footprints of solar magnetic field lines open into the heliosphere, coronal holes provide a critical measure of the structure and evolution of these lines. Using a combination of \textit{Solar and Heliospheric Observatory / Extreme ultraviolet Imaging Telescope} (SOHO/EIT), \textit{Solar Dynamics Observatory / Atmospheric Imaging Assembly} (SDO/AIA) and \textit{Solar Terrestrial Relations Observatory / Extreme Ultraviolet Imager}\break(STEREO/EUVI A/B) extreme ultraviolet (EUV) observations spanning 1996--2015 (nearly two solar cycles), coronal holes are automatically detected and characterized. Coronal hole area distributions show distinct behavior in latitude, defining the domain of polar and low-latitude coronal holes. The northern and southern polar regions show a clear asymmetry, with a lag between hemispheres in the appearance and disappearance of polar coronal holes.
\end{abstract}

%

\end{opening}

%

\section{Introduction}

Coronal holes are the observational signatures of regions of open solar magnetic field. Early observations were conducted in extreme ultraviolet (EUV) wavelengths \citep{1968IAUS...35..411T}. Magnetic field lines within the solar corona can either connect back down to the solar surface or extend outward into the heliosphere, denoting closed and open field respectively. In regions of open magnetic field, coronal plasma is free to follow these field lines outward, creating the solar wind and a plasma density depletion at the field footpoints. Appearing as dark regions in X-ray and EUV images, coronal holes correspond with reduced emission from solar plasma localized to open magnetic field footpoints \citep{2009SSRv..144..383W}. \cite{2009LRSP....6....3C} discusses a complete review of coronal holes in exhaustive detail.

The Sun's open magnetic field and coronal hole structure varies alongside the overall solar activity cycle, running through a magnetic polarity swap every 11 years. The connection between coronal hole activity and open magnetic flux throughout the solar cycle has been considered by several studies \citep{2002SoPh..211...31H,2009SSRv..144..383W}. \cite{2002SoPh..211...31H} assessed the properties of polar coronal holes throughout solar Cycles 22 and 23. Using a series of He {\sc i} 10830 \AA~spectroheliograms, polar coronal holes were identified over the period 1989 September to 2002 March, and their area and enclosed magnetic flux was measured. They found that polar coronal holes throughout this period develop initially at lower latitudes of about 50-60 degrees, and extend to the poles within three subsequent rotations. These polar coronal holes then have a lifetime observed between 8.3 and 8.7 years. An asymmetry was observed in the timing of the initial appearance and evolution of coronal holes at each pole, with a difference of several months between poles. This mirrors earlier observations by \cite{1984SoPh...92..109W}, who found a lag of 9 and 6 months between polar coronal hole appearances between cycles 19-20 and 20-21, respectively. The relative areal extent of each polar coronal hole can also show an asymmetry from one solar cycle to the next \citep{1978SoPh...56..161B,1980SoPh...65..229S}. \cite{2002JGRA..107.1302W} explored in detail through observation and modeling the relationship between active region evolution and consequent coronal hole distribution. In particular, they noted the evolution of polar coronal holes as a consequence of remaining flux transported from lower latitudes. More recent work by \cite{2013ApJ...765..146M} explored the hemispheric asymmetry of photospheric magnetism throughout cycles 23 and the early stages of cycle 24. While He {\sc i} 10830 \AA~observations of coronal holes can provide accurate results in polar regions, there is discrepancy at lower-latitudes \citep{1983SoPh...87...47K,2003SoPh..212..165S,2005SoPh..226....3M}.

Despite the relatively long history of observations of the solar activity cycle, continuous and full-disk magnetic field and solar corona observations extend back only to cover the previous two cycles. The \textit{Solar and Heliospheric Observatory} (SOHO) began this observational campaign from 1995 through 2011, with results from the \textit{Solar Dynamics Observatory} (SDO) picking up since May 2010. \cite{2014ApJ...792...12M} recently conducted a comprehensive analysis of observations by these spacecraft, showing that distributions of EUV bright points and structures of surface magnetic elements, characterized by the magnetic range of influence (MRoI), both migrate from high latitudes ($\pm$55 degrees) toward the equator in a time period of 19 years. In addition, the evolution of low-latitude coronal holes and records of coronal green line emissions also overlaps with this trend. \cite{2010ApJ...716..693R} explored this apparent extended cycle as observed from coronal emission in simulations of green line emission and \textit{Extreme ultraviolet Imaging Telescope} (EIT) EUV observations. They suggested that these overlapping band structures, rather than indicating an extended cycle, are merely remnants of reconnection between low latitude and polar flux from the current cycle. Observing the progress of the activity bands formed by these very different features helps decipher the underlying solar magnetism in a cycle structure, and can help to shed additional light on these processes.

Among these features, coronal holes map the footprints of open magnetic field at locations that vary with the solar cycle. \cite{2014ApJ...792...12M} showed that low-latitude coronal holes migrate towards the equator as the cycle progresses, whereas during solar minimum holes exist primarily in polar regions, where EUV bright points and coronal green line emission is sparse. These are the areas of crucial importance in the solar magnetic activity cycle. The polar region is usually dominated by magnetic field of one polarity, and each cycle begins with the reversal of this polarity at the end of the solar minimum, when meridional convective flow has recycled the field to the poles and flux cancellation takes place \citep{2014ApJ...780....5U}. Existing models of global solar magnetic field often make the point to focus on comparison of computed open field models with the observed coronal holes at the poles, or with the total heliospheric open magnetic flux \citep{2002SoPh..207..291M, 2002SoPh..209..287M, 2010JGRA..115.9112Y}. On the other hand, much effort is also needed to examine coronal holes evolving across the latitudes. Tracking open magnetic flux on the Sun following coronal hole evolution from the poles to the low-latitudes adds a valuable piece to the puzzle of the solar magnetic activity cycle.

The continuous full-disk coronal and magnetic field observations in the past two decades provide the opportunity to study the evolution of coronal holes and open magnetic flux in solar Cycles 23 and 24. \cite{2014ApJ...783..142L} developed an automated technique to detect persistent coronal holes and characterize their properties using a database of full-disk EUV images obtained by several spacecraft instruments over a time span from 1996 May to 2013 January. In this present study, we use the same technique and the same database which however is further extended to August 2014. The extended database covers a significant portion of the current solar Cycle 24, and hence provides an opportunity to compare coronal hole properties between the past Cycle 23 and this special new cycle, which, after an unexpectedly deep minimum, slowly started around 2010, and rose to a low maximum in early 2014 \citep{2015NatCo...6E6491M}.

\cite{2014ApJ...783..142L} have also measured coronal hole areas, and the unsigned and signed total flux measured in these holes, distinguishing polar regions from low-latitudes with an arbitrary division at $\pm$65 degrees. This paper presents a more careful and detailed examination of the latitude dependence of coronal hole properties, where a contrast between the consecutive two cycles is apparent. A brief review of the database and technique is presented in Section 2. Section 3 presents the latitude dependence of coronal hole properties and their variations from the last cycle to the current cycle, followed by conclusions and discussions in Section 4.


\section{Methodology and Data}

\begin{table}
\caption{Data coverage for each instrument source.}
\label{table:table1}
\begin{tabular}{rrrrrr}
\hline
 & & \multicolumn{2}{c}{Start}  & \multicolumn{2}{c}{End} \\ 
\cline{3-4} \cline{5-6} \\ 
Source & Observable & CR & Date & CR & Date \\
\hline
SOHO/EIT				& EUV 195\AA & 1909.96 & 1996 05 31 & 2105.27 & 2010 12 31\\
SDO/AIA	 				& EUV 193\AA & 2096.76 & 2010 05 13 & 2153.92 & 2014 08 19\\
STEREO/EUVI			 	& EUV 195\AA & 2096.76 & 2010 05 13 & 2153.92 & 2014 08 19\\
WSO Harmonics	& $B_r$ & 1893.00 & 1995 02 23 & 2140.00 & 2013 08 04\\
Synoptic SOHO/MDI & $B_r$ & 1911.00 & 1996 06 28 & 2104.00 & 2010 11 26\\
Synoptic SDO/HMI & $B_r$ & 2096.00 & 2010 04 22 & 2156.00 & 2014 10 14
\end{tabular}
\end{table}

EIT \citep{1995SoPh..162..291D}, an instrument onboard SOHO, has provided fourteen years of nearly continuous EUV observations over the span of solar cycle 23. SDO was launched in 2010, continuing the role of a provider of synoptic EUV observations through the EUV telescope \textit{Atmospheric Imaging Assembly} \citep[AIA,][]{2012SoPh..275...17L}. Having launched a few years prior, the twin spacecraft \textit{Solar Terrestrial Relations Observatory} (STEREO) A and B have swept out a significant portion of vantage points on the far-side of the Sun. Employing the \textit{Extreme Ultraviolet Imager} \citep[EUVI,][]{2008SSRv..136...67H} on each of the STEREO spacecraft, additional viewpoints are available in the EUV. The particular combination of EUV data from SDO/AIA and STEREO/EUVI A/B allows for a unique observational opportunity. As the STEREO spacecraft have swept out in their orbits, expanded heliographic longitude coverage in EUV has been made possible. Additionally, polar EUV coverage was expanded by the more staggered $B$-angle for each of the spacecraft in use. Joint observations by the three spacecraft complement one another to provide better coverage of the polar areas, important regions contributing to the evolution of solar open magnetic flux. The joint AIA/EUVI coverage makes possible continuous, consistent, and nearly-full solar surface observations of coronal hole boundaries over solar Cycle 24. Table~\ref{table:table1} displays the relevant datasets used in this study, and the corresponding availability ranges. Using these data ranges, coronal hole boundaries are characterized over one and one half solar activity cycles, and compared between Cycles 23 and 24.

\begin{figure*}[htbp]
\begin{center}
\includegraphics[width=12cm]{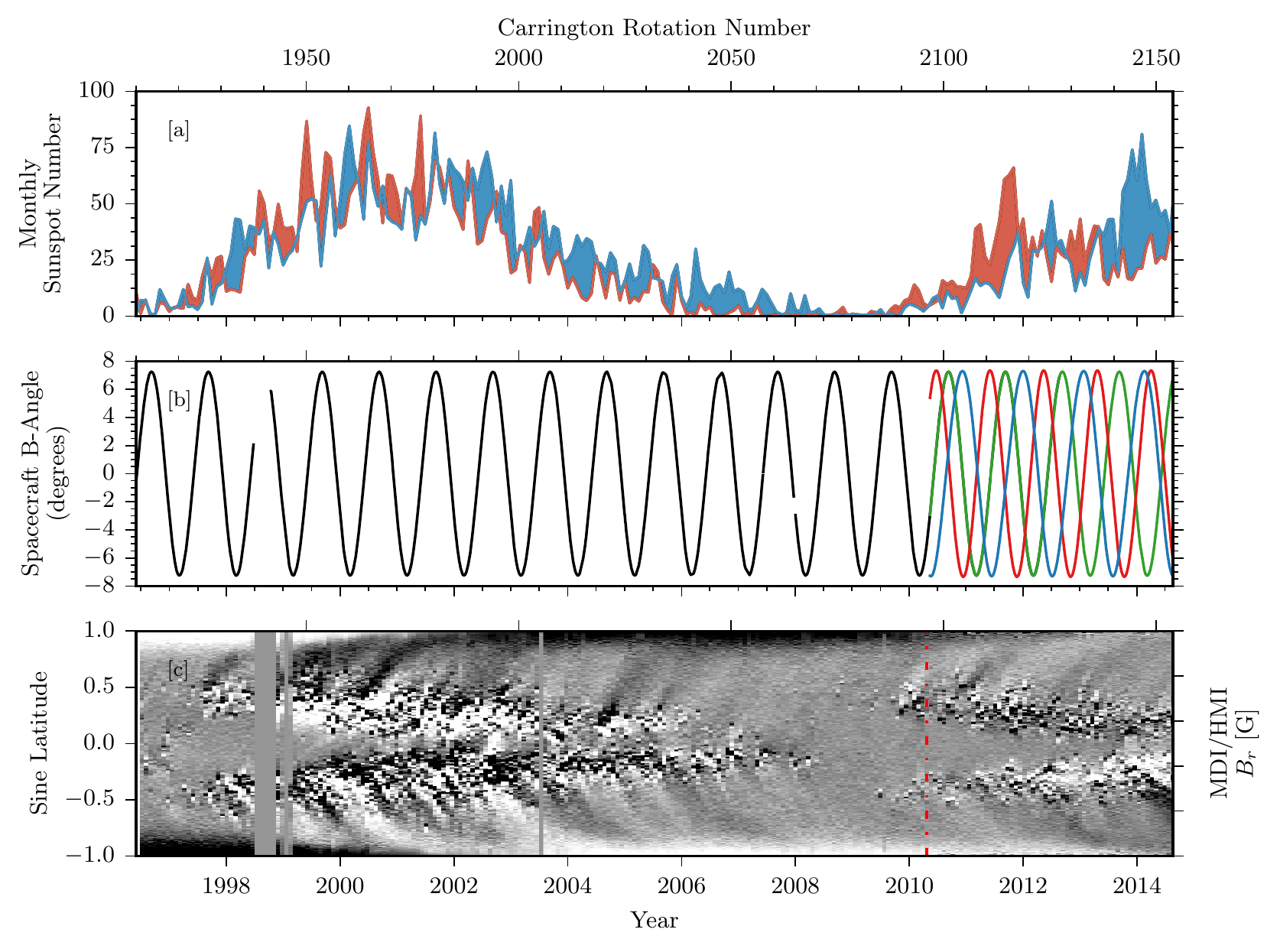}
\caption{[a] Hemispheric sunspot number for northern and southern hemispheres in red and blue, respectively. (WDC-SILSO, Royal Observatory of Belgium, Brussels.) [b] Spacecraft solar $B$-angle over the course of observations in Table~\ref{table:table1}. SOHO/EIT (black), SDO/AIA (green), STEREO/EUVI A (red), and STEREO/EUVI B (blue) angles are displayed. [c] Butterfly diagram (latitude-time profile) of mean radial magnetic field strength, scaled between $\pm$ 5 G. Measurements from SOHO/MDI transition to those from SDO/HMI at the vertical dot-dashed red line in early 2010.}
\label{fig:scft_bangle}
\end{center}
\end{figure*}

To provide context of the solar cycle dependence of coronal hole properties, Figure~\ref{fig:scft_bangle} displays the observational spacecraft $B$-angle in comparison with position in the solar activity cycle. Some of this data was featured in the prior study. However, here this range is extended to cover through August 2014 \citep[][Figure 13]{2014ApJ...783..142L}. The upper panel displays the monthly hemispheric sunspot number (WDC-SILSO, Royal Observatory of Belgium, Brussels), with the northern and southern sunspot counts in red and blue, respectively, and with the dominant pole shade filling between. The middle panel displays the spacecraft $B$-angle for SOHO/EIT (black), SDO/AIA (green), STEREO/EUVI A (red), and STEREO/EUVI B (blue). Using the single vantage point of SOHO/EIT, our polar observations are limited by this angle. This will become apparent in subsequent figures. Rather than attempting a convoluted methodology to correct or estimate for this lack of coverage, results are displayed unmodified. With the combination of SDO/AIA and STEREO/EUVI A/B, this polar coverage gap is greatly reduced. In comparing with our gauge of solar activity cycle, note that our EIT observations span over the entire first solar cycle under consideration. Our improved polar coverage from STEREO begins conveniently just as the next solar cycle gets underway. The bottom panel displays a butterfly diagram of mean radial magnetic field strength, taken from measurements from the \textit{Michelson Doppler Imager} \citep[MDI,][]{1995SoPh..162..129S} and the \textit{Helioseismic and Magnetic Imager} \citep[HMI,][]{2012SoPh..275..207S}, with a vertical dot-dashed red line to indicate the transition. This data is scaled to $\pm$ 5 G to highlight trends in dominant magnetic field sign throughout the solar cycle.

With these data, techniques have been developed and documented for the automated detection of coronal hole boundaries, detailed in literature \citep{2014ApJ...783..142L}. In brief, each frame of EUV data is analyzed via an instrument-independent intensity thresholding method to mark out coronal hole boundaries. Boolean maps of suspected coronal hole regions are generated and mapped into a Carrington Equal Area (CEA) projection, and then segmented and catalogued into individual regions through a watershed method. For our SOHO-era (solar activity Cycle 23) data, this process is repeated for each time-step, with half of the solar surface not visible. Full rotation maps are built by summing these individual EIT hole maps over each Carrington rotation, and then flattening into a boolean map. For the SDO-era (solar activity Cycle 24) data, thresholded projections are made individually for each of our three datasets (SDO/AIA and STEREO/EUVI A/B), and then summed and flattened into one resulting boolean map.

This initial mapping detects both coronal holes and filament channels, which both appear as dark features in EUV. Here we use the underlying magnetic flux density information to distinguish the two. Filament channels stride magnetic polarity inversion lines, encompassing roughly equal distributions of positive and negative flux. In the encompassed area of a coronal hole, a single magnetic polarity dominates \citep{2009LRSP....6....3C}. Therefore, we calculate the skew of magnetic flux density in each candidate coronal hole region, using magnetograms from SOHO/MDI in conjunction with SOHO/EIT, or SDO/HMI with SDO/AIA and STEREO/EUVI A/B. Regions with a relatively small skew are considered to be filament channels and are therefore `sieved' out, leaving behind a boolean CEA map of coronal hole pixel locations.

For the current application of this routine for long-duration coronal hole evolution, the EUV data cadence is as follows. SOHO/EIT data are assimilated daily, with a resulting daily coronal hole map. These daily maps are then summed and flattened to provide an upper estimate over an entire solar rotation, to account for missing far-side data. SDO/AIA and STEREO/EUVI A/B data are gathered at a cadence of 12 hours, providing two full-surface coronal hole maps per day. Note that with the expanded longitudinal and polar coverage, these maps are considered individually, without the requirement of summing over an entire solar rotation.

For more details and intermediate results on this process, see \cite{2014ApJ...783..142L}. For the sake of brevity, this code and existing methodology will be referred to as the Global Automated Coronal Hole Detection routine (GACHD). Here the techniques have been further refined, and applied to the complete data range from 31 May 1996 to 19 August 2014, the furthest data available until the recent communication loss with one of the STEREO spacecraft. These data cover the entire past Cycle 23 as well as a large portion of the current solar Cycle 24.

\section{Coronal Hole Properties in Solar Cycles 23 and 24}

Using this dataset of coronal hole boundaries, spanning over one and one half solar cycles in length, a few interesting quantities can be considered. These include the area as well as the signed and unsigned magnetic flux enclosed by coronal hole boundaries. Rather than considering just the polar coronal hole properties, our data extend across all latitudes to characterize the latitude-dependence of these properties and their evolution.

In this section, we will first present the time-latitude maps of coronal hole properties. The total magnetic flux in the holes is then presented, comparing the polar regions with low-latitudes, and the past Cycle 23 with the current cycle. For the following set of figures, there are two notes of importance. Figures~\ref{fig:lprof_cyc_ssn}, \ref{fig:ch_area_cyc}, \ref{fig:ch_flux_cyc}, and \ref{fig:ch_sflux_cyc} display quantities over the full time span from May 1996 to August 2014, illustrating the solar cycle dependence of the measured properties. Figures~\ref{fig:lprof_sdo_ssn}, \ref{fig:ch_area_sdo}, \ref{fig:ch_flux_sdo}, and \ref{fig:ch_sflux_sdo} provide a `zoomed-in' display of identical quantities from 2010 May onward, over the SDO and STEREO-era data for solar Cycle 24. Over all of these plots, panel a displays the monthly hemispheric sunspot number in red and blue, for the northern and southern sunspot counts, respectively. This panel will lead some of the following figures for an easily markable solar activity cycle cross-reference tool.

It should be noted that a partial overlap with the results in \cite{2014ApJ...783..142L} exists within Figures~\ref{fig:ch_area_cyc}, \ref{fig:ch_flux_cyc}, and \ref{fig:ch_sflux_cyc}. However, results contained here use an alternative latitudinal definition of polar regions. Additional data extends the range of results to cover a more significant fraction of Cycle 24.

\subsection{Latitude Coronal Hole Profiles}

\begin{figure*}[htbp]
\begin{center}
\includegraphics[width=12cm]{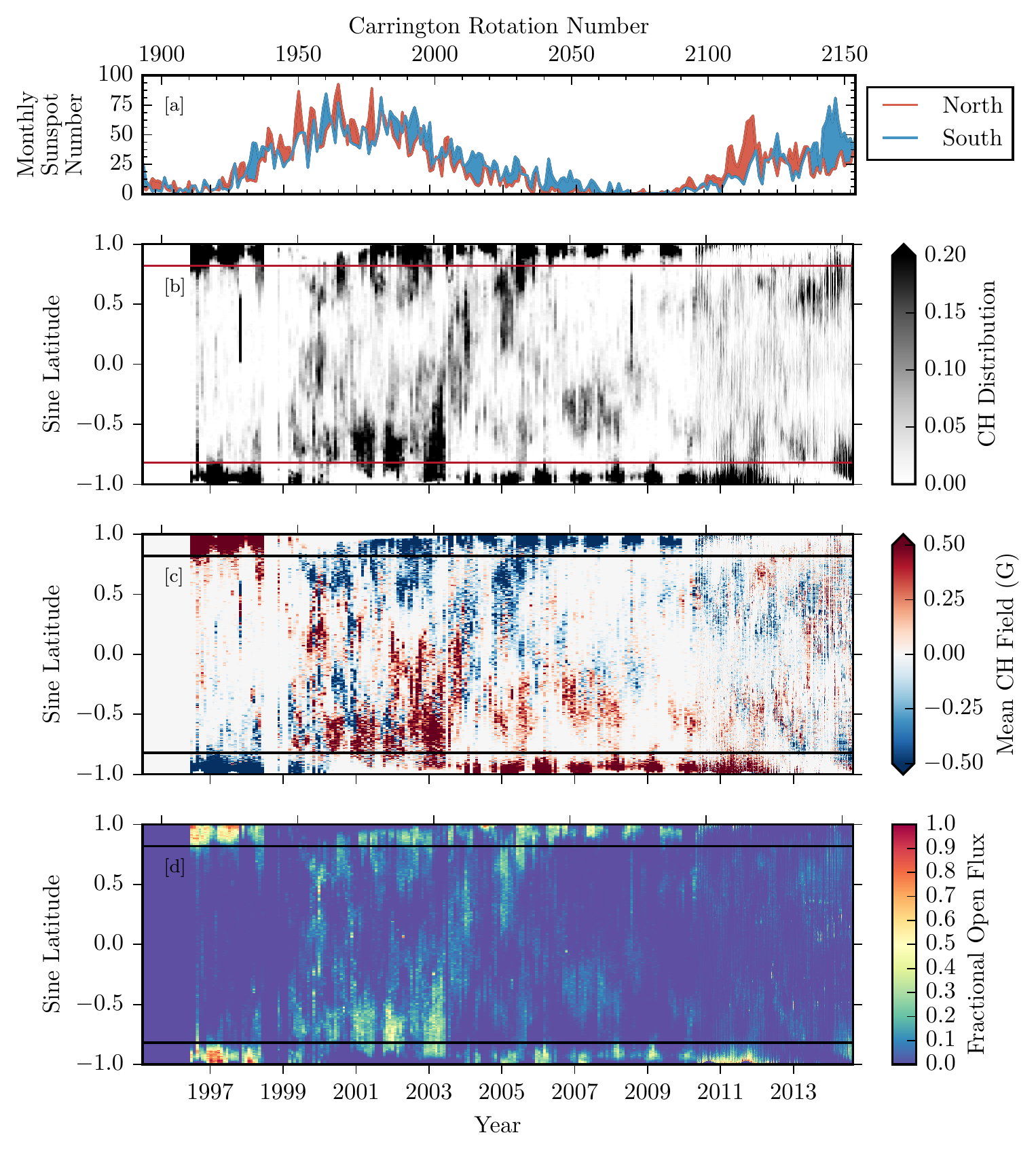}
\caption{For the time period 31 May 1996 to 19 August 2014, the following are shown: [a] Monthly hemispheric sunspot number for the northern (red) and southern (blue) hemispheres (WDC-SILSO, Royal Observatory of Belgium, Brussels). [b] Coronal hole latitude profile of distribution, integrated over longitude. Horizontal red lines are marked at $\pm$55 degrees latitude, to distinguish polar and low-latitude zones. [c] Latitude profile of coronal hole dominant polarity, integrated over longitudes. [d] Latitude profile of coronal hole-enclosed magnetic flux, integrated over longitudes.}
\label{fig:lprof_cyc_ssn}
\end{center}
\end{figure*}

Consider Figure~\ref{fig:lprof_cyc_ssn}, displaying the full latitude dependence of our coronal hole quantities over the full timespan of our datasets. Of immediate note, distributions of the coronal hole area (panel b), and the mean magnetic field (panel c) and total unsigned magnetic flux (panel d) in the holes all exhibit distinctive evolution patterns near $\pm$55-degrees latitude, which are marked by two horizontal lines in the figures.

Other authors have arrived at slightly different limiting latitudes for polar coronal hole boundaries. \cite{2002JGRA..107.1302W} considered the formation and evolution of polar field as beginning at far lower latitudes. In this formalism, remnants of active region flux are transported towards the poles, where shearing and diffusion work to reduce the nonaxisymmetric component of this field. The remaining field component continues towards the poles, establishing a dominant polarity. In this manner, coronal holes begin their formation in lower latitudes, $\pm$30-60 degrees, before establishing themselves above $\pm$60 degrees, with boundaries somewhat maintained by a balance between meridional flow and diffusion.

Similarly, observations of polar coronal holes in He {\sc i} 10830 \AA~spectroheliograms by \cite{2002SoPh..211...31H} found that most held positions above $\pm$60 degrees, with some exceptions. Taking the mean polar coronal hole latitude, \cite{2002SoPh..211...31H} found dips in the northern and southern poles to 58-59 degrees and 54-55 degrees, respectively, during the Cycle 22 to 23 minimum. Therefore, in the following analysis, $\pm$55 degrees latitude is chosen as the division between polar and low-latitude regions. While lower than the $\pm$60 degree boundary, this provides a clear distinction of polar regions, without misattributing open magnetic flux. Note that this distinction of polar regions differs from the previous study by \cite{2014ApJ...783..142L}, which used $\pm$65 degrees latitude as a separator. Appendix~\ref{sec:appendix} works through a more detailed analysis of the effects of shifting this polar latitude boundary, with further justification for this boundary choice. Note that the selection here of a polar latitude is useful for comparative purposes.

These particular latitudes, $\pm$55-degrees, have been noted previously as a demarcation of certain other aspects of solar activity, coincidentally. \cite{1996Sci...272.1300T}, using the \textit{Global Oscillation Network Group} (GONG) set of observations, considered the internal rotation rate of the Sun, from the surface deep into the convection zone. A slice through the convection zone reveals the internal structure of the differential rotation, beneath the photospheric surface flow. At $\pm$55-degrees latitude, there is a distinct change in the behavior of these subsurface flows. \cite{2011sswh.book...39S}, in an effort to sketch out a few theories of internal solar dynamo processes, devised a simple model. Starting with a uniform poloidal field, a differential rotation profile from observations was applied. This resulted in a shearing of the field, which can eventually lead to a buildup of azimuthal field and an instability. From this simple model, the azimuthal field builds as $\partial_t B_\phi \sim \sin 2 \lambda (1+ 1.51 \sin^2 \lambda)$. As a function of latitude, the maximum shearing occurs around $\pm$55 degrees. \cite{1983ApJ...270..288S} finds this same latitude as a peak of differential rotation shearing. This latitude was further revealed as a zone of interest by \cite{2014ApJ...792...12M,2015NatCo...6E6491M}, who found a scarcity of EUV bright points above this latitude, as well as a distinct evolution pattern of the magnetic range of influence (MRoI) divided by this latitude. While the coronal hole polar boundary is determined by properties of the polar field itself, several other quantities show variation nearby.

Panel b of Figure~\ref{fig:lprof_cyc_ssn} displays for each timeframe and bin of sine latitude the fraction of longitude bins that are characterized as coronal holes, with the color mapping truncated at 0.2 for display purposes. Note that this fraction varies from 0 (no coronal holes at any longitudes at that particular latitude) to 1.0 (complete coronal hole coverage at every longitude for that particular latitude). From 1996 to 1998, and from 2006 through 2012, coronal holes are concentrated in the polar regions above $\pm$55-degrees. Note here that the polar hole extensions from 1996 to 1998 extend further equatorward, pushing towards our boundaries at $\pm$55 degrees. Apparent annual variation in the polar hole area is not a physical result, but rather a consequence of the solar $B$-angle modulation of the SOHO spacecraft (see Figure~\ref{fig:scft_bangle}). The single vantage point of these observations results in alternating reduced polar coverage. From 2010, with the combination of the AIA/EUVI suite of data, staggered spacecraft $B$-angles allow for more comprehensive polar measurements. During the solar maximum from 1999 to 2003, and then from 2012 through 2014, coronal holes extend to low-latitudes below $\pm$55-degrees, and are diminished in polar regions.

Figure~\ref{fig:lprof_cyc_ssn} panel c displays the mean magnetic field strength encompassed by coronal hole boundaries, averaged over all longitude bins for each bin of sine latitude. Using synoptic magnetograms from SOHO/MDI and SDO/HMI, coronal hole enclosed magnetic flux (signed and unsigned) is integrated over all longitudes. To avoid the contribution of noise for both signed and unsigned flux calculations, instrument-specific noise levels are used to remove magnetic flux density measurements below this noise threshold. For MDI synoptic chart measurements of radial magnetic flux density this noise level is measured as $\pm$5.0~G, while for similar synoptic measurements from HMI this noise level is measured as $\pm$2.3 G \citep{2012SoPh..279..295L}. While the range of integrated values ranges from $-8.4$ to 14.3 G, here the map is displayed truncated at $\pm$0.5~G to illustrate the often-weak polarity dominance at lower latitudes. The resulting map displays the distribution of the dominant polarity of coronal hole-enclosed magnetic flux and its evolution over the solar cycle. These distributions of dominant coronal hole polarity exhibit a hemispheric pattern over the course of the solar cycle. In the rise of the previous cycle before 1999, the northern polar coronal hole distribution is dominated by positive magnetic field, with the southern polar coronal hole dominated by negative field. At lower latitudes, this same general polarity trend is observed, albeit at a much lower strength, usually below 1~G, and the trend is not as spatially uniform across latitudes and time. During the 2001-2003 solar activity maximum, the hemispheric field reversal is observed in the swap of the polar coronal hole dominant polarity, persisting until 2012. For the current cycle, the reversal of the dominant polarity in polar regions begins in 2012 and 2013 for the northern and southern hemisphere, respectively. However, the resulting polar coronal hole coverage and polarity dominance have been very weak in comparison with the prior cycle. In lower latitudes during the current cycle, the distribution of coronal hole polarity dominance is much weaker in strength and spatially mixed.

Figure~\ref{fig:lprof_cyc_ssn} panel d displays the fractional unsigned open flux, collected by summing all coronal-hole-encompassed-flux over longitude bins for each time and sine-latitude bin. An identical noise cutoff of the synoptic magnetogram is used for this calculation to avoid including instrument noise. Note that the range of calculated open flux ranges from 0 to 4.7$\times 10^{20}$ Mx, with displayed data represented as a fraction of the maximum measured. With a sine-latitude coordinate system, broken into 1440 pixel latitude bins, each pixel represents an equal solar surface area. This allows for comparison of the relative strength of open flux sources at polar and low-latitude regions. From 1997--1999 the polar coronal holes contain the majority of the open magnetic flux. As the poles reverse as solar activity ramps up, larger concentrations of open flux appear at lower latitudes between 1999 and 2003. Past the maximum of Cycle 23, this pattern quickly returns to larger concentrations of open flux in the poles, until the next reversal around 2013. In the southern pole, the concentration of coronal hole flux persists for the extended deep minimum until the next reversal around 2013. This particular cycle from 2009 onward has shown unique behavior in the early drop-off of the northern polar coronal hole signature. This in turn has resulted in almost nonexistent open-flux from the northern polar region.

In concluding this portion of the analysis, we note that \cite{2014ApJ...792...12M,2015NatCo...6E6491M} have also reported the solar-cycle dependence of coronal holes observed in much the same time range. However, their study only captured coronal holes with area greater than 20,000~Mm$^2$ at low latitudes (below $\pm$55 degree). By employing joint observations by AIA/EUVI, our study not only provides a more complete coverage of polar holes from May 2010, but also allows the assessment of polar hole measurements by EIT in the past cycle through comparison of measurements by different instruments, as will be further discussed in Sections 3.2 and 3.3.

\subsection{Coronal Hole Area}

\begin{figure*}[htbp]
\begin{center}
\includegraphics[width=12cm]{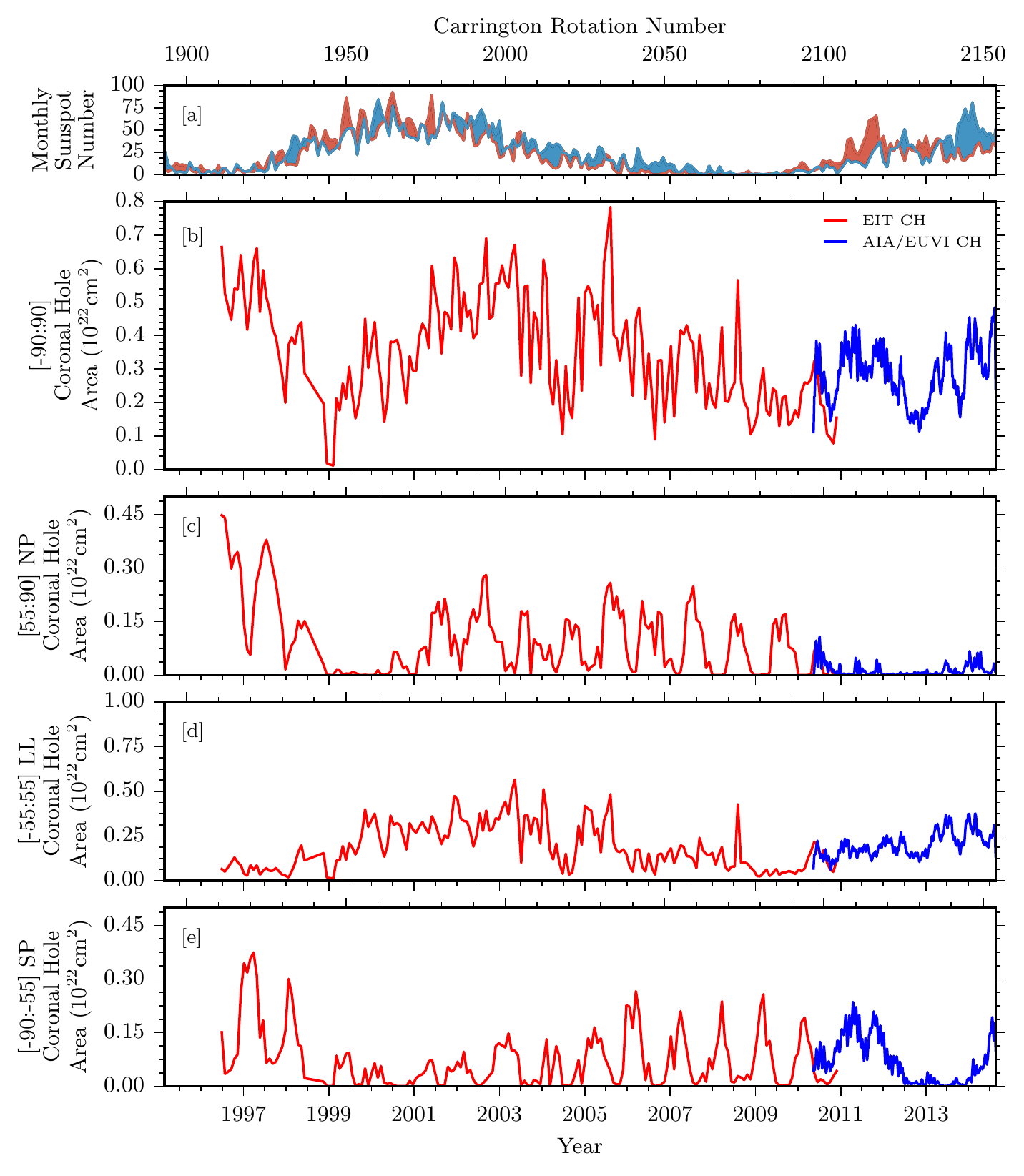}
\caption{Comparisons of coronal hole area and associated quantities over the entire data span. [a] Hemispheric sunspot number for northern and southern hemispheres in red and blue, respectively. (WDC-SILSO, Royal Observatory of Belgium, Brussels.) [b] Total coronal hole area, for some of our data sources. EIT and AIA/EUVI coronal hole boundaries are used to compute the coronal hole area shown in red and blue, respectively. [c-e] Coronal hole area computed for northern polar, low latitudinal, and southern polar regions.}
\label{fig:ch_area_cyc}
\end{center}
\end{figure*}

Figure~\ref{fig:ch_area_cyc} breaks down the latitude dependence of the coronal hole areas in Figure~\ref{fig:lprof_cyc_ssn} into the distinct ranges of Northern Polar (NP; 55:90 degrees), Low Latitudinal (LL; -55:55 degrees), and Southern Polar (SP; -90:-55 degrees). Results from EIT and the combination of results from AIA/EUVI are displayed in red and blue, respectively. Note here that EIT measurements of area are integrated over each solar rotation, providing a measurement of area within observable regions. The brief overlap of EIT and AIA/EUVI observations in the second half of 2010 demonstrates that, as also discussed in \cite{2014ApJ...783..142L}, the joint AIA/EUVI observational coverage allows better estimates of polar coronal hole areas and fluxes.

Figure~\ref{fig:ch_area_cyc} panel b displays results for the entire range of latitudes, from $-90$:$90$ degrees. The total area of coronal holes over the course of this study varies between 1--7$\times 10^{21}$ cm$^2$, or from 2\% to 12\% of the total solar surface area. Figure~\ref{fig:ch_area_cyc} panels c-e display the latitude zones under consideration, where we can note a few things. The northern coronal hole has dropped from 2011 onward, hovering around zero area, and has never recovered to the pre-2009 level. The southern pole area was still rising from 2011, and suffered a drop in the middle of 2012, but recovered from 2014 onward. In comparing with the previous cycle drop in polar coronal hole area from 1999-2001, this behavior is highly asymmetric, reflecting asymmetries in sunspot activity visible in Figure~\ref{fig:scft_bangle}.

In both cycles, low-latitude coronal holes cover a large area during the solar maximum, and consequently are a source of a significant amount of total open flux on the Sun during the solar maximum. In the current Cycle 24, the low-latitude coronal hole area is notably smaller than in the last cycle.

\subsection{Unsigned Magnetic Flux}

\begin{figure*}[htbp]
\begin{center}
\includegraphics[width=12cm]{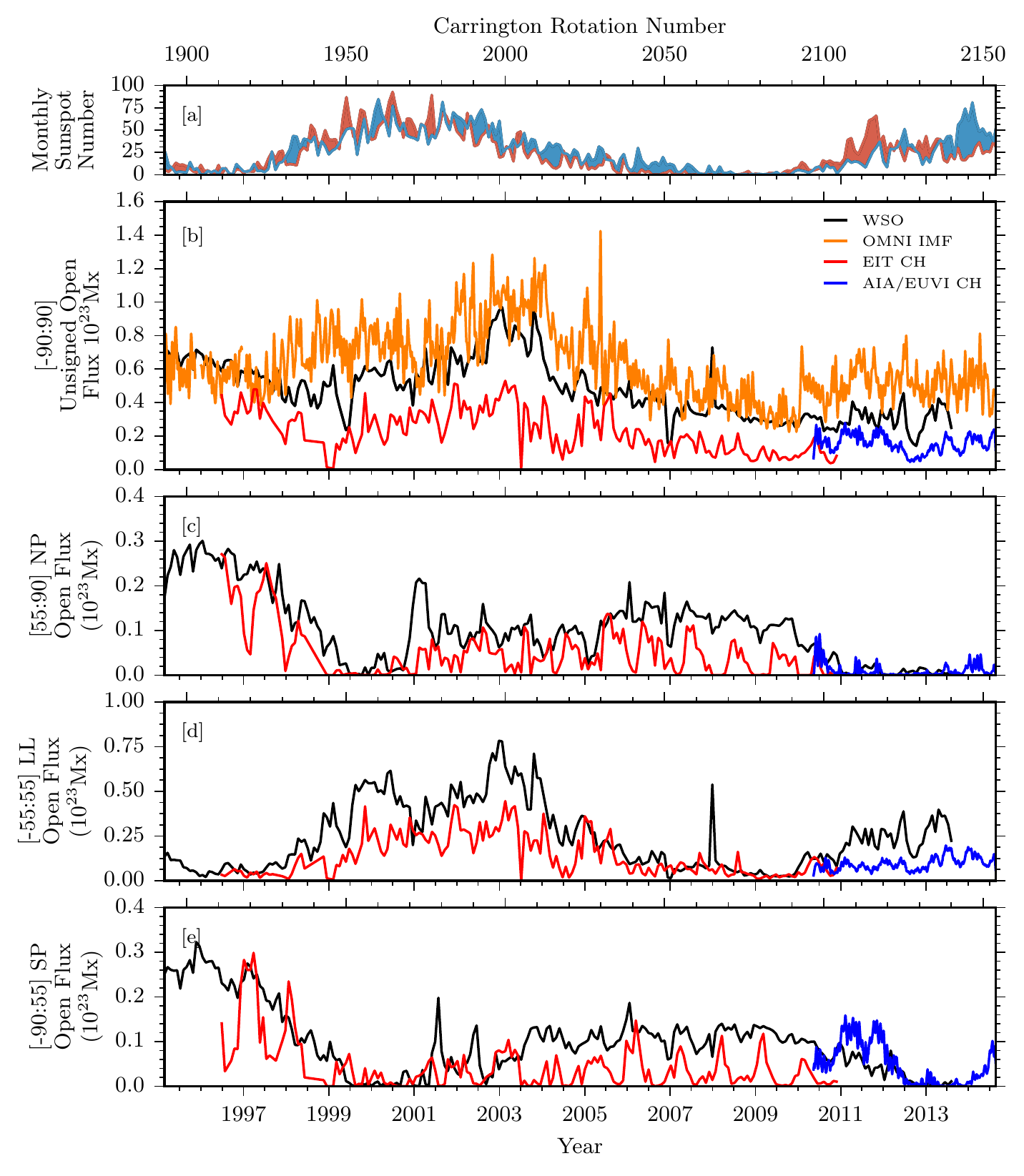}
\caption{Comparisons of unsigned photospheric flux and associated quantities over the entire data span. [a] Hemispheric sunspot number for northern and southern hemispheres in red and blue, respectively. (WDC-SILSO, Royal Observatory of Belgium, Brussels.) [b] Total unsigned open magnetic flux, for some of our data sources. WSO data in black is computed from total open field boundaries. EIT and AIA/EUVI coronal hole boundaries are used to compute the total enclosed magnetic flux in red and blue, respectively. OMNI $|B_r(r=1$AU$)|$ is used to compute an equivalent open magnetic flux, displayed in orange. [c-e] Open magnetic flux computed for northern polar, low latitudinal, and southern polar regions.}
\label{fig:ch_flux_cyc}
\end{center}
\end{figure*}

In a similar manner, Figure~\ref{fig:ch_flux_cyc} breaks down the latitude dependence of the unsigned open magnetic flux encompassed by our coronal hole boundaries. Figure~\ref{fig:ch_flux_cyc} panel b displays the total unsigned open flux, with several datasets in place. EIT and AIA/EUVI results for coronal hole-encompassed unsigned flux are plotted in red and blue, respectively. These direct measurements are compared with the open flux calculated by the potential field source surface (PFSS) model \citep{1969SoPh....6..442S, 1992ApJ...392..310W}. This model calculates the magnetic potential field using spherical harmonics of photospheric radial magnetic field determinations from the \textit{Wilcox Solar Observatory} (WSO), and with the source-surface defined at 2.5$R_\odot$, where field lines are assumed open. An array of field lines are traced down from this source-surface to their photospheric origin, and are mapped out as the boundaries of open magnetic field. Magnetic fluxes are then computed by integrating the photospheric radial magnetic field within these open footpoint boundaries. Note that while WSO observations are used as a boundary condition, the calculated open flux is a PFSS model output using this observational input. For the purposes of comparison with other model results, this dataset may be referred to as `observational,' referring to the observational basis of the model input. The resulting unsigned open magnetic flux from this PFSS model is displayed in black. 

The measured open flux from the EIT coronal holes undershoots the values predicted by the PFSS model, partially due to a contrast issue with SOHO/EIT \citep{2014ApJ...783..142L}. This issue is most prevalent in lower-latitude regions during periods of higher solar activity. AIA/EUVI measurements reveal additional open flux as compared with EIT alone during the overlapping period in the second half of 2010, as a result of more complete coverage of the polar regions by the AIA/EUVI dataset. This additional flux sets the AIA/EUVI dataset more in line with the PFSS model.

Finally, the OMNI dataset is used to compute an equivalent open magnetic flux. In-situ (Lagrange point L1) measurements of the $B_x$ component (Geocentric Solar Ecliptic; GSE) of the magnetic field extend backward for decades through a cross-spacecraft calibrated dataset \citep{2005JGRA..110.2104K}. The OMNI data are obtained from the GSFC/SPDF OMNIWeb interface at \url{http://omniweb.gsfc.nasa.gov}. For this study we focus only on data for the previous two decades. If we make the assumption that our long-term interplanetary magnetic field is uniform and radial, the equivalent OMNI unsigned open magnetic flux can be computed as $\Phi_{\mathrm{OMNI}} = 4 \pi R_{L1}^2 |B_x|$ \citep{2002ESASP.508..507L}. During solar minimum periods, the OMNI data and PFSS model are in better agreement than in periods of higher solar activity. 

Figures~\ref{fig:ch_flux_cyc} panels c-e display the coronal hole unsigned open flux for the latitude zones demarcated by $\pm$55 degrees latitude. Similar evolution patterns are observed with regard to cyclic variation and latitudinal dependence, as compared with measurements of coronal hole areas. In the polar regions, the lack of coverage during the EIT-era dataset is apparent, as a consequence of $B$-angle variation. However, the peak values of EIT coronal hole area and unsigned open flux provide an upper bound on this value for each pole during optimal angling. Considering the upper bounds of the EIT data, good agreement is found with the PFSS open flux in polar regions. This trend continues with AIA/EUVI data, albeit with more complete coverage of the polar regions. Such comparisons allow us to state that the magnetic open flux in polar regions computed with the PFSS model is generally consistent with direct measurements in both solar cycles.

In terms of the solar cycle dependence of open flux in the poles, for solar Cycle 23, the unsigned open flux in the northern and southern poles evolved in a mostly symmetric manner. But an asymmetry presented itself with solar Cycle 24, just as with the evolution of the coronal hole areas. Once again, this is a consequence of asymmetric sunspot emergence. For this cycle, the northern polar unsigned open flux drops off to nearly zero around 2011, and continues to hang there. Slight signs of the beginning of a recovery are apparent in 2014, but with no clear recovery trend. The southern pole open flux drops to zero in 2013, two years later than the northern poles, and then begins to rise once again.

Consideration of the low-latitude open flux reveals the discrepancy during periods of maximum solar activity, when low-latitude coronal holes border close to larger concentrations of magnetic flux.

\subsection{Signed Magnetic Flux}

\begin{figure*}[htbp]
\begin{center}
\includegraphics[width=12cm]{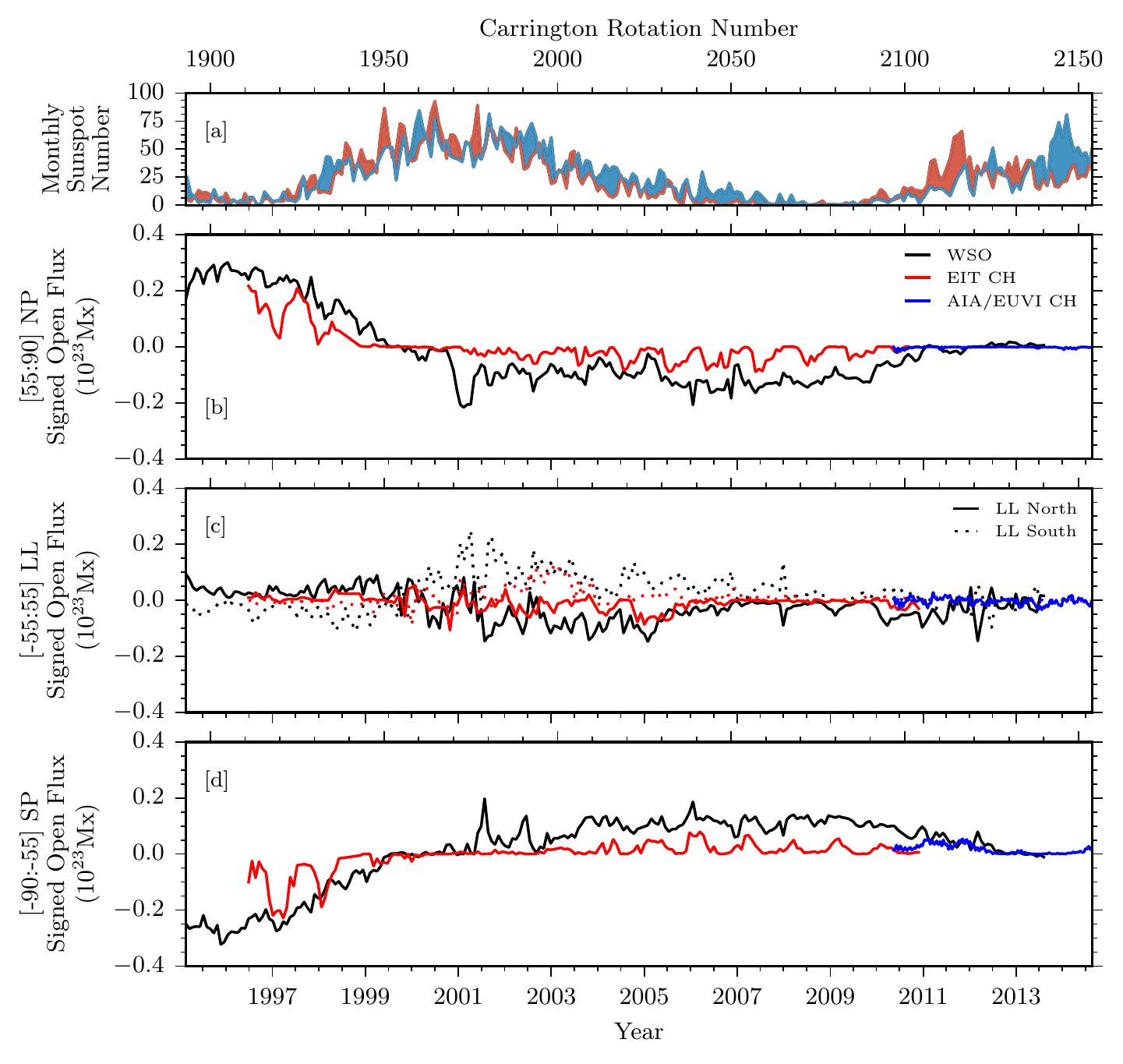}
\caption{Comparisons of signed photospheric flux and associated quantities over the entire data span. [a] Hemispheric sunspot number for northern and southern hemispheres in red and blue, respectively. (WDC-SILSO, Royal Observatory of Belgium, Brussels.) [b-d] Signed open magnetic flux, for some of our data sources, for the northern polar, low-latitudinal (northern and southern portions in solid and dotted styles), and southern polar regions. WSO data in black is computed from total open field boundaries. EIT and AIA/EUVI coronal hole boundaries are used to compute the total enclosed magnetic flux shown in red and blue, respectively.}
\label{fig:ch_sflux_cyc}
\end{center}
\end{figure*}

Figure~\ref{fig:ch_sflux_cyc} panels b-d display the signed open magnetic flux from each of the latitude zones demarcated by $\pm$55 degrees. Considering the northern and southern low latitude regions of Figure~\ref{fig:ch_sflux_cyc} panel c, the signed open flux undergoes a few variations throughout the cycle. During solar minima, the signed open fluxes are relatively weaker, due to a reduced low-latitude area during these phases. As the areas ramp up during Cycle 23, each low-latitude hemisphere begins to exhibit a dominant signed polarity, positive in the south and negative in the north. This behavior ramps down during the decline to solar minimum between Cycles 23 and 24. As Cycle 24 begins to ramp up, observations of low-latitude hemispheric signed flux remain relatively weak.

For the northern and southern polar regions, shown in Figure~\ref{fig:ch_sflux_cyc} panels b-d, the polarity dominance is more apparent. During the lull in solar activity around 1997 before Cycle 23 kicks off, each polar region is strongly dominated. The northern pole shows a strong dominance of positive magnetic polarity, while the southern pole shows a strong negative dominance. Both our EIT coronal hole and WSO computed open field measurements agree. As solar maximum for Cycle 23 is reached, this polarity dominance drops off, and each pole becomes neutral around 2000. Throughout this next bout of minimal activity between Cycles 23 and 24, a different story emerges. Here, the WSO computed open field shows a much weaker polarity dominance, this time of the opposite sign as expected. This is also reflected in a weaker polar unsigned flux. Our EIT measurements of coronal hole open flux show an even weaker polarity dominance. This stands in stark contrast with the heavy polarity dominance of the previous minimum. As Cycle 24 reaches something resembling a peak around 2014, once again both poles have reverted to a neutral state. Here our AIA/EUVI observations fall more in line with the computed WSO open field measurements.

\subsection{A Close Look at Cycle 24}

\begin{figure*}[htbp]
\begin{center}
\includegraphics[width=12cm]{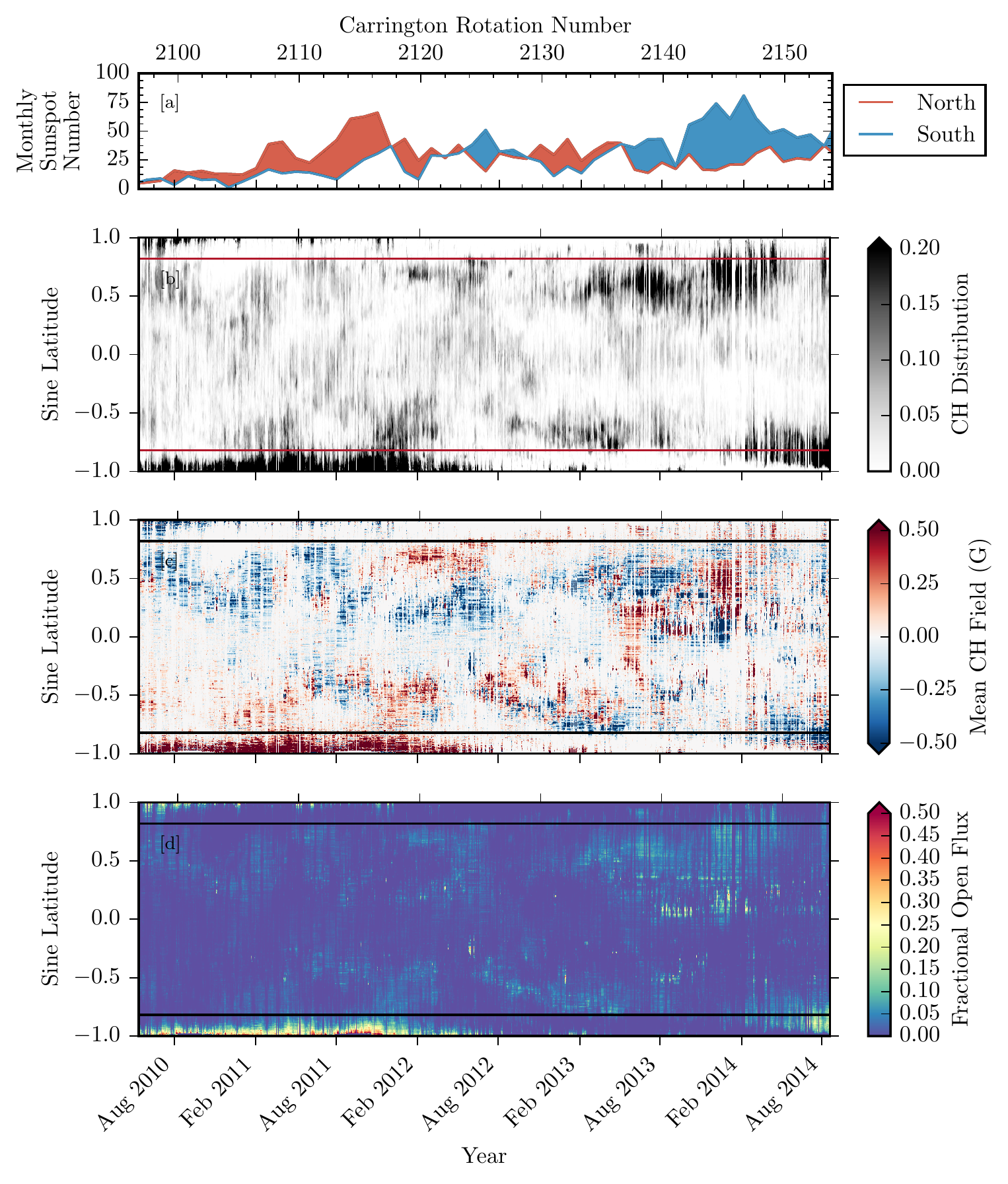}
\caption{For the time period 13 May 2010 -- August 19 2014 the following are shown: [a] Monthly hemispheric sunspot number for the northern (red) and southern (blue) hemispheres. (WDC-SILSO, Royal Observatory of Belgium, Brussels.) [b] Coronal hole latitude profile of distribution, integrated over longitude. Horizontal red lines are marked at $\pm$55 degrees latitude, to distinguish polar and low-latitude zones. [c] Latitude profile of coronal hole dominant polarity, integrated over longitudes. [d] Latitude profile of coronal hole enclosed magnetic flux, integrated over longitudes.}
\label{fig:lprof_sdo_ssn}
\end{center}
\end{figure*}

\begin{figure*}[htbp]
\begin{center}
\includegraphics[width=12cm]{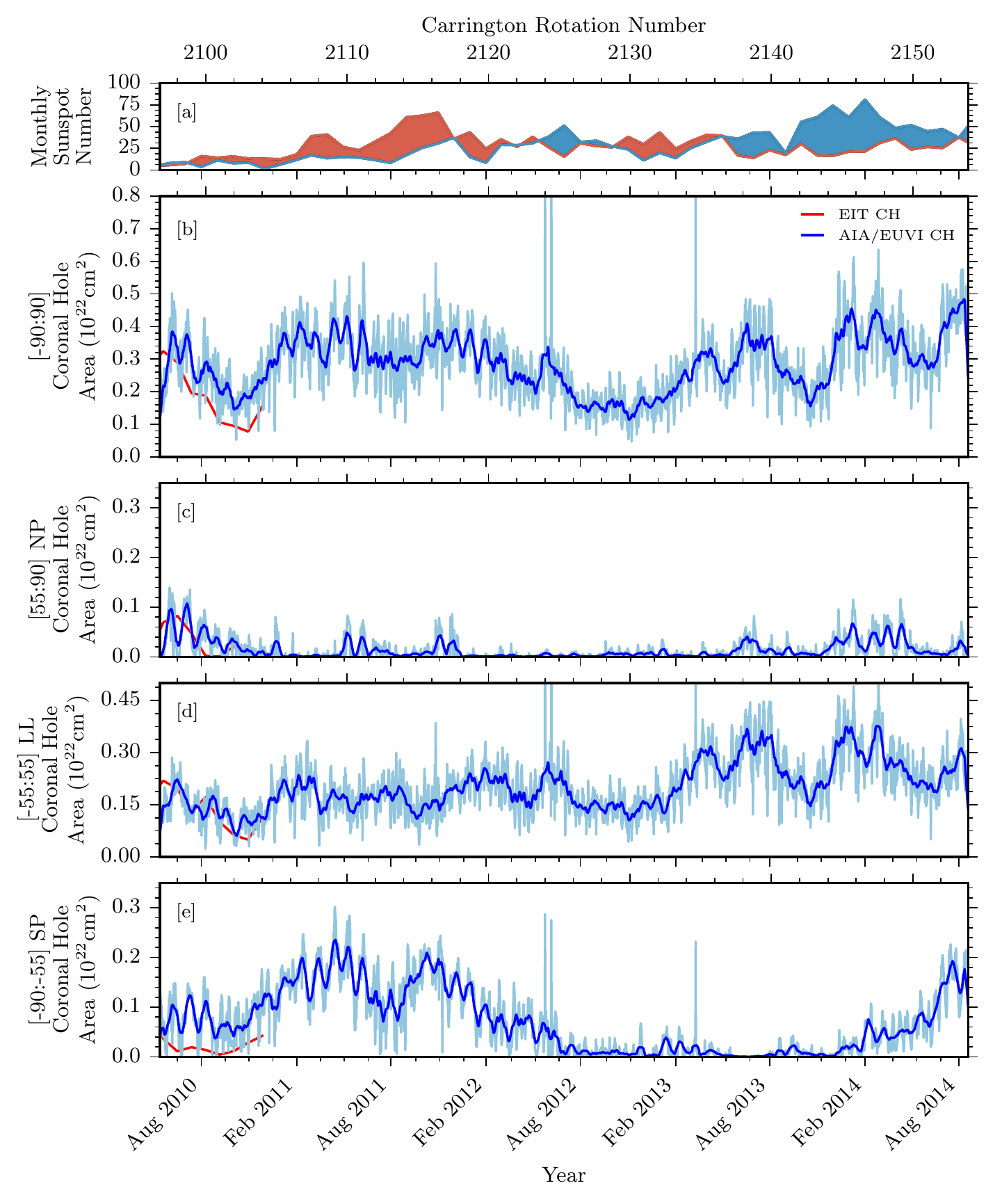}
\caption{Comparisons of coronal hole area and associated quantities over the SDO-era data span. [a] Hemispheric sunspot number for northern and southern hemispheres in red and blue, respectively. (WDC-SILSO, Royal Observatory of Belgium, Brussels.) [b] Total coronal hole area, for some of our data sources. EIT and AIA/EUVI coronal hole boundaries are used to compute the coronal hole area in red and blue, respectively. AIA/EUVI raw data are displayed in light-blue, with a 27-day running average displayed in dark-blue. [c-e] Coronal hole area computed for northern polar, low latitudinal, and southern polar regions.}
\label{fig:ch_area_sdo}
\end{center}
\end{figure*}

\begin{figure*}[htbp]
\begin{center}
\includegraphics[width=12cm]{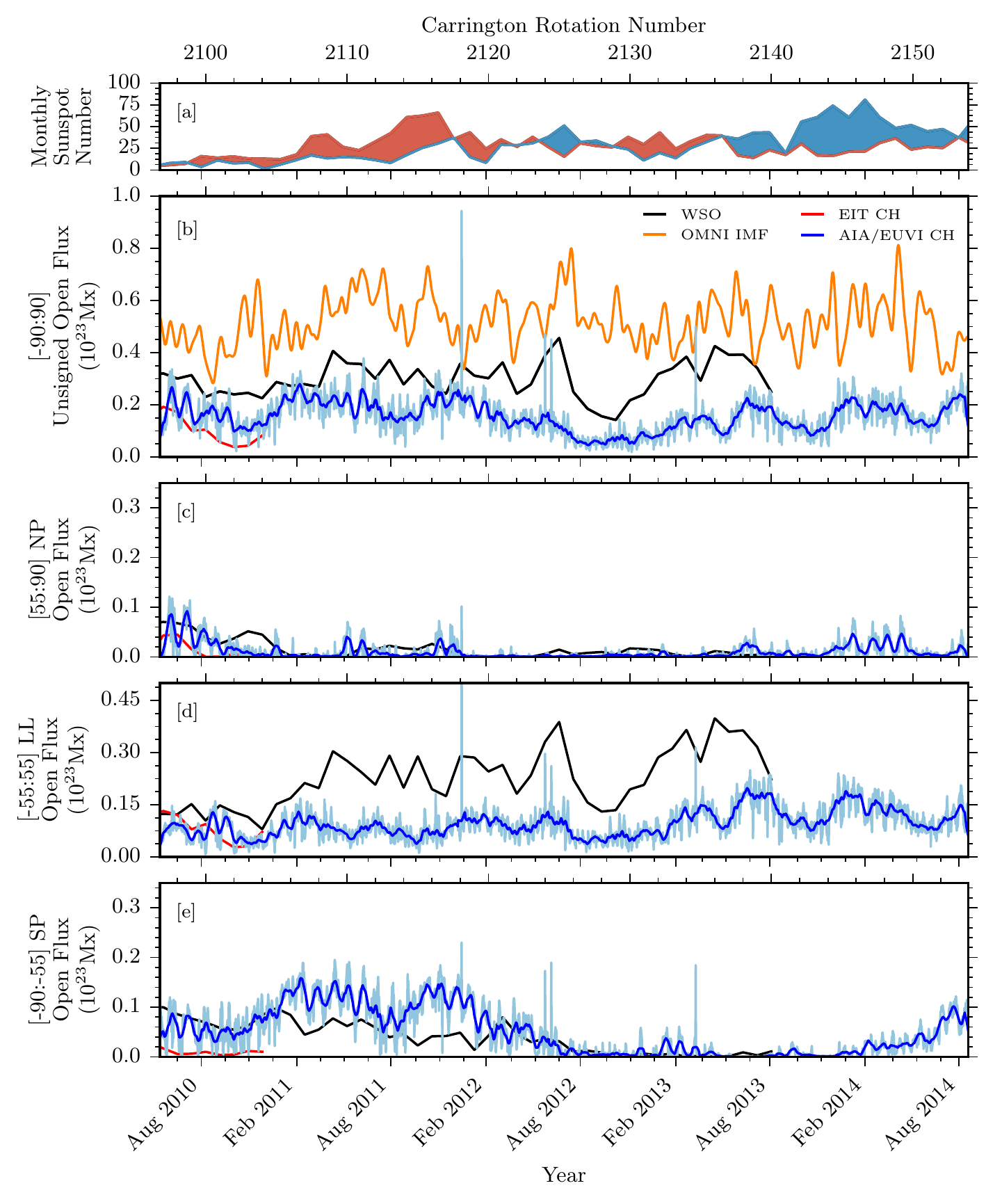}
\caption{Unsigned photospheric flux and associated quantities over the SDO-era data span. [a] Hemispheric sunspot number for northern and southern hemispheres in red and blue, respectively. (WDC-SILSO, Royal Observatory of Belgium, Brussels.) [b] Total unsigned open magnetic flux. WSO computed open flux is displayed in black. EIT and AIA/EUVI coronal hole boundaries are used to compute the total enclosed magnetic flux in red and blue, respectively. AIA/EUVI raw data are displayed in light-blue, with a 27-day average displayed in dark-blue. OMNI $|B_r(r=1$AU$)|$ is used to compute an equivalent open magnetic flux (orange). [c-e] Open magnetic flux computed for northern polar, low latitudinal, and southern polar regions.}
\label{fig:ch_flux_sdo}
\end{center}
\end{figure*}

\begin{figure*}[htbp]
\begin{center}
\includegraphics[width=12cm]{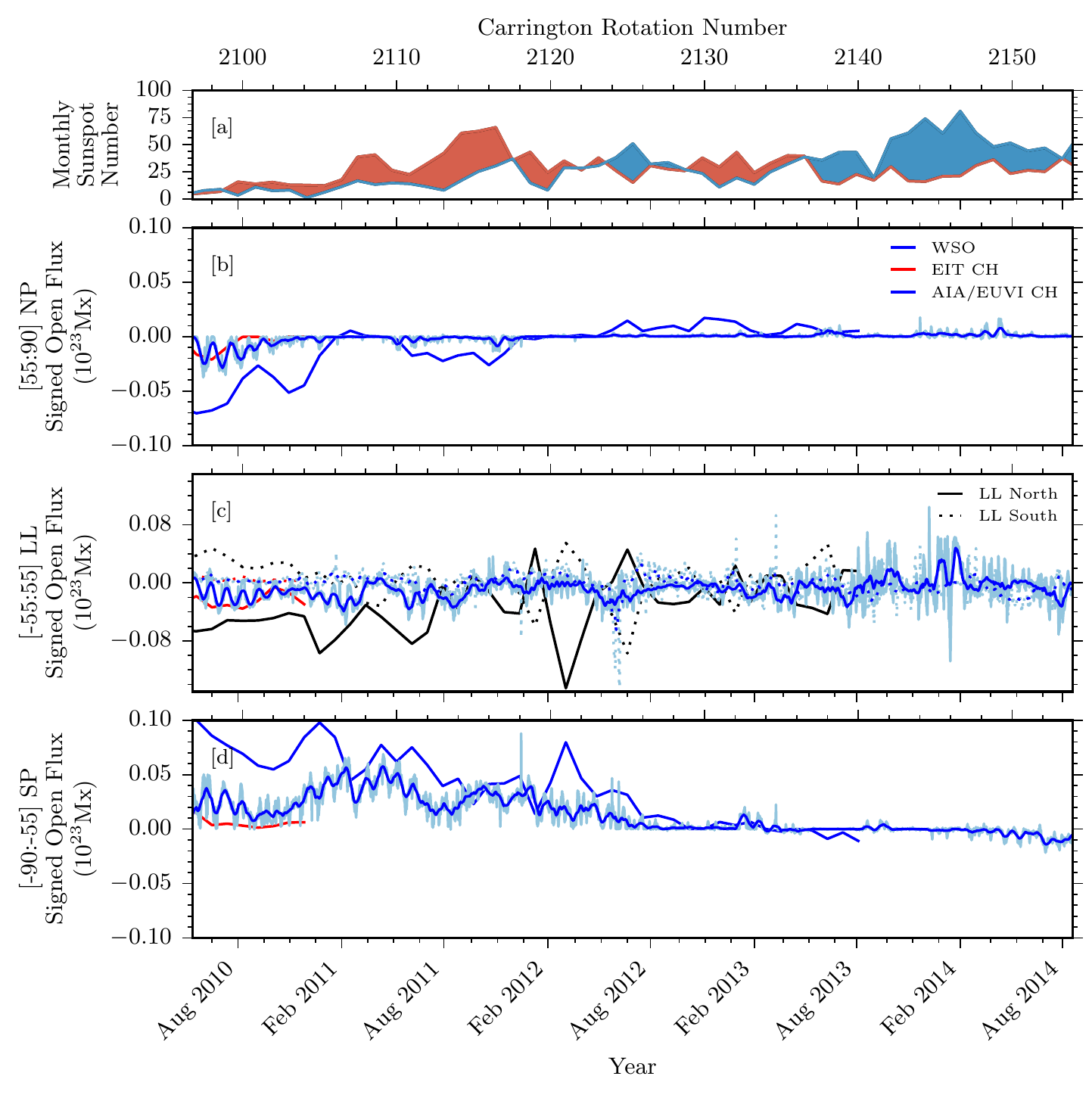}
\caption{Comparisons of signed photospheric flux and associated quantities over the SDO-era data span. [a] Hemispheric sunspot number for northern and southern hemispheres in red and blue, respectively. (WDC-SILSO, Royal Observatory of Belgium, Brussels.) [b-d] Signed open magnetic flux, for some of our data sources, for the northern polar, low-latitudinal (northern and southern portions in solid and dotted styles), and southern polar regions. WSO data in black is computed from total open field boundaries. EIT and AIA/EUVI coronal hole boundaries are used to compute the total enclosed magnetic flux in red and blue, respectively. AIA/EUVI raw data are displayed in light-blue, with a 27-day running average displayed in dark-blue.}
\label{fig:ch_sflux_sdo}\end{center}
\end{figure*}

To better illustrate details for solar cycle 24, we consider data only from 13 May 2010 -- 19 August 2014, Carrington rotations 2096--2154. Here, many of the same trends and conclusions are evident, with some new details visible with scaling. Figures~\ref{fig:lprof_sdo_ssn}, \ref{fig:ch_area_sdo}, \ref{fig:ch_flux_sdo}, and \ref{fig:ch_sflux_sdo} display identical values to the larger dataset just discussed, but with a retracted viewpoint to only cover the range of data available from AIA/EUVI. Figure~\ref{fig:lprof_sdo_ssn} gives maps showing latitude distribution of the coronal hole, mean magnetic field and total unsigned flux in the holes. Figures~\ref{fig:ch_area_sdo}, \ref{fig:ch_flux_sdo}, and \ref{fig:ch_sflux_sdo} represent the measured coronal hole area, unsigned open magnetic flux, and signed open magnetic flux, respectively. These plots are identical in labeling to those spanning the entire dataset, with the addition of AIA/EUVI raw data displayed in the lighter shade of blue, compared with a 27-day running average in darker blue.

Figure~\ref{fig:lprof_sdo_ssn} panels b-d displays the SDO-era data available using our coronal hole detection routine. While our SOHO-era coronal hole data was gathered at 24-hour cadence, the results were stacked and binned over a full solar rotation. This methodology allows for the calculation of an upper-bound of coronal hole area and enclosed unsigned magnetic flux, capturing the location of each Earth-facing coronal hole location throughout that rotation. Our SDO-era methodology is slightly different, with the inclusion of STEREO/EUVI data. With the inclusion of EUVI data, far-side (opposite Earth) coronal hole observations are possible. This allows for our coronal hole mappings to be presented at full 12-hour cadence, without binning. This is immediately apparent with the coronal hole area latitude profiles in Figure~\ref{fig:lprof_sdo_ssn} panel b. In May 2010, the beginning of our dataset, a regular gap in the data appears, coinciding with the solar rotation rate. At the beginning of our SDO-era data, 13 May 2010, the STEREO spacecraft had achieved an orbital separation of roughly 70-degrees ahead and behind, each with respect to Earth. This left roughly a 40-degree range of longitudes that were unobservable in EUV. As the STEREO spacecraft continued in their orbital separation, this gap continued to shrink. With near-limb data truncated from each data source, higher frequency gaps appear further in the dataset as large holes rotate through these small blind-spots.

SDO-era latitude profiles of coronal hole area and signed/unsigned flux displayed in Figure~\ref{fig:lprof_sdo_ssn} reveal a few items are of note. From May 2010 until January 2014, the latitude demarcation of $\pm$55 degrees continues to divide our polar and low-latitude coronal holes. From January 2014 until August 2014, two large coronal holes develop on the upper edge of our low-latitude region, then drift and extend beyond $\pm$55 degrees, with the northern polar extension of this hole developing much weaker and diminishing earlier than that found in the southern pole. Referring back to Figure~\ref{fig:lprof_cyc_ssn} panel b, this same pattern of expansion into the pole is apparent with EIT data from 2001 until 2003. Here, however, this expansion occurs with much greater symmetry and nearly in phase; both the northern and southern extensions into the polar regions evolve and persist. The asymmetric coronal hole distribution within polar regions in Cycle 24 is apparent in the total coronal hole area displayed in Figure~\ref{fig:ch_area_sdo}.

The mean coronal hole field in Figure~\ref{fig:lprof_sdo_ssn} shows the dominant polarity of each of our major coronal hole distributions. From May 2010 until August 2012 the northern polar coronal hole distribution is dominated by negative magnetic flux, with the southern polar coronal holes dominated by positive flux. Note that the northern polar coronal hole distribution has fallen off in area and resulting signed/unsigned flux as of February 2011, with lingering traces thereafter until completely vanishing in February 2012. Once again, this particular lead-in to a polar polarity inversion behaves in a much more asymmetric manner than the reversal observed with SOHO-era data in mid-1999.

The unsigned open flux latitude distribution in Figure~\ref{fig:lprof_sdo_ssn} panel d results from previous noted trends in the coronal hole area latitude profile. Here we see that the southern polar coronal hole distribution dominates the density of unsigned open magnetic flux, with smaller distributions at the northern pole and lower latitudes. The measured total flux is plotted in Figure~\ref{fig:ch_flux_sdo}. As described in \cite{2014ApJ...783..142L}, as a whole the AIA/EUVI measured total flux is much closer to the model computed total flux, but there are variations in the latitude dependent comparison. The AIA/EUVI measured flux is comparable or higher than the model computed flux within polar regions. However, at lower latitudes the computed flux overshoots the measured flux.

\section{Conclusions and Discussions}

Through the use of SOHO/EIT, SDO/AIA, and STEREO/EUVI A$/$B EUV data from May 1996 until August 2014, coronal hole boundaries have been tracked over the entirety of solar Cycle 23, and a good portion of Cycle 24. This extensive range allows for detailed analysis of coronal hole evolution in all latitudes, and comparison of the magnetic open flux, which is directly measured from the coronal holes, from the past cycle to the current cycle. The addition of STEREO/EUVI data allows for improved polar coverage in the current cycle, and also provides a reference to assess the polar hole and open flux measurements by EIT in the past cycle. A few trends in this data are apparent.

These data show that coronal holes, and the signed and unsigned magnetic flux measured inside the holes, exhibit distinct evolution patterns as a function of latitude near the polar regions. Also observed is an asymmetry in the evolution of coronal holes in the northern and southern polar regions, consistent with observations by \cite{2014ApJ...792...12M}. This asymmetry manifests in both the rising of polar coronal hole coverage during solar Cycle 23, and more distinctively in the decline and rise of polar coronal holes in solar Cycle 24. By tracking the evolution of coronal holes across all latitudes, we see the appearance of large dominant-polarity coronal hole structures at lower-latitudes, migrating towards the polar regions. While these low-latitude signatures appear at roughly identical times, their push into the polar regions occurs at differing speeds, resulting in the northern polar coronal hole establishing itself much earlier. 

When the two cycles are compared in their early stages, the current cycle has a similar coronal hole extent and total open flux. Most significantly, in Cycle 24, the polar flux reversal exhibits a stark asymmetry compared with the last cycle. The northern polar holes have dropped off from 2009, two years ahead of the southern polar holes, and the open flux there has remained at a very low level since then, whereas the southern polar holes have recovered from the drop-off in 2013 with the open flux rising again in 2014. It remains to be seen how the differences of cycle 24 will affect the peak distribution of coronal holes and open magnetic flux during the declining phase of Cycle 24.

Finally, the observed coronal holes, directly measured open flux in these holes, and their latitude dependence have been compared with the calculation by the PFSS model using the synoptic WSO data. It is shown that, on the one hand, this model reproduces polar coronal holes and open flux in general agreement with observations in both cycles. On the other hand, confirming the results in our earlier study \citep{2014ApJ...783..142L}, there is a clear discrepancy at low latitudes, which contributes significantly to the total open flux, particularly during the solar maximum. The potential field model appears to produce more open flux than observed at the low latitudes in both cycles, even though these regions are presumably better captured by the AIA/EUVI instruments in the current cycle. Therefore, such comparison indicates the importance of the latitude-dependent evolution of coronal hole boundaries, which provide an observational constraint to help advance models of the Sun's global magnetic field.

%
\appendix

\section{Polar Latitude Determination}\label{sec:appendix}

In the preceding analysis, a polar boundary latitude was chosen at $\pm$55 degrees. While the latitude-time profiles display variation over a continuum of latitude bins, a polar boundary value was necessary for quantifying coronal hole area and enclosed flux across specified ranges of latitudes. This appendix works through an analysis of the effects of the shifting polar boundary, providing justification for the choice of boundary.

\begin{figure*}[htbp]
\begin{center}
\includegraphics[width=12cm]{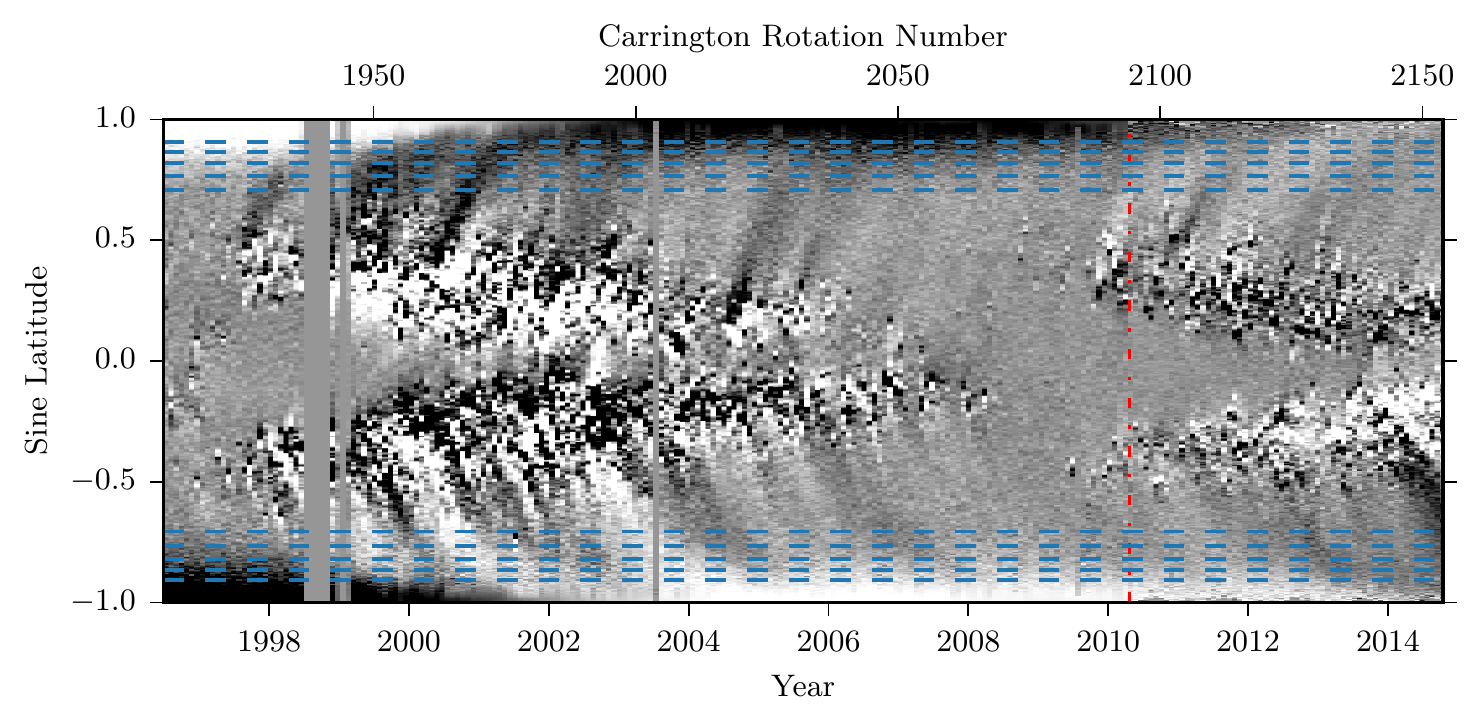}
\caption{Butterfly diagram (latitude-time profile) of mean radial magnetic field strength, scaled between $\pm$ 5 G. Measurements from SOHO/MDI transition to those from SDO/HMI at the vertical dot-dashed red line. Latitude boundaries at $\pm$\{45, 50, 55, 60, 65\} degrees are marked by blue dotted lines.}
\label{fig:bfly_varlat}\end{center}
\end{figure*}

Figure~\ref{fig:bfly_varlat} displays the mean radial magnetic field strength, capped at $\pm$ 5 G. Measurements from SOHO/MDI are plotted until 22 April 2010 (marked with a dot-dashed red line), after which SDO/HMI data are mapped. To study a range of latitude values, latitudes of $\pm$\{45, 50, 55, 60, 65\} degrees were chosen as boundaries for study, to give a wide range of values for comparison. These latitude values are marked here by blue dotted lines. During the period from 2001--2014, the distribution of strong polar magnetic flux remains above $\approx$60 degrees latitude. However, a difference in this distribution appears for available data prior to 2001. Here, the distribution of polar flux appears to dip further equatorward.

\begin{figure*}[htbp]
\begin{center}
\includegraphics[width=12cm]{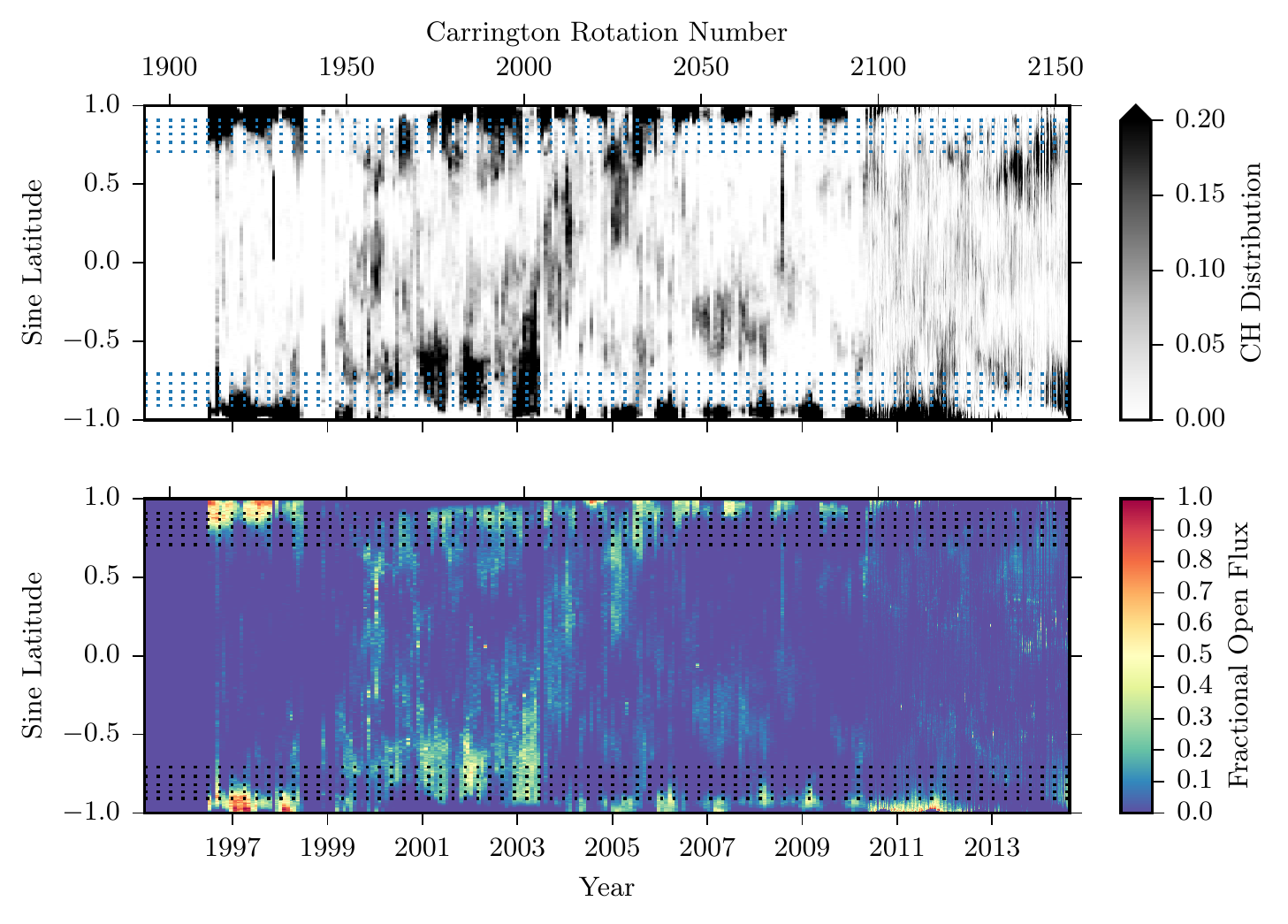}
\caption{[upper] Coronal hole latitude profile of distribution, integrated over longitude. [lower] Latitude profile of coronal hole enclosed magnetic flux, integrated over longitudes. Latitude boundaries of $\pm$\{45, 50, 55, 60, 65\} degrees are marked by blue dashed lines in both panels.}
\label{fig:lprof_cyc_ssn_varlat}\end{center}
\end{figure*}

This difference in extent of polar magnetic field manifests in changes to the distribution of polar open magnetic field. Figure~\ref{fig:lprof_cyc_ssn_varlat} maps out the latitudinal distribution of coronal hole boundaries and the flux enclosed therein through the span of observed data. For comparison, latitudes of $\pm$\{45, 50, 55, 60, 65\} degrees are plotted in dotted styles in both panels. Note that the remainder of the data here is identical to associated panels in Figure~\ref{fig:lprof_cyc_ssn}. From the period 2001--2014, the resulting coronal hole distribution remains relatively above $\pm$ 60 degrees latitude. However, in the time spans 1996--1998 and 2011--2012, extensions of coronal holes dip below this, encroaching closer to $\pm$ 55 degrees. The lower panel displays the corresponding coronal hole-enclosed unsigned open magnetic flux. These extensions to lower latitudes carry a significant corresponding amount of open magnetic flux.

\begin{figure*}[htbp]
\begin{center}
\includegraphics[width=12cm]{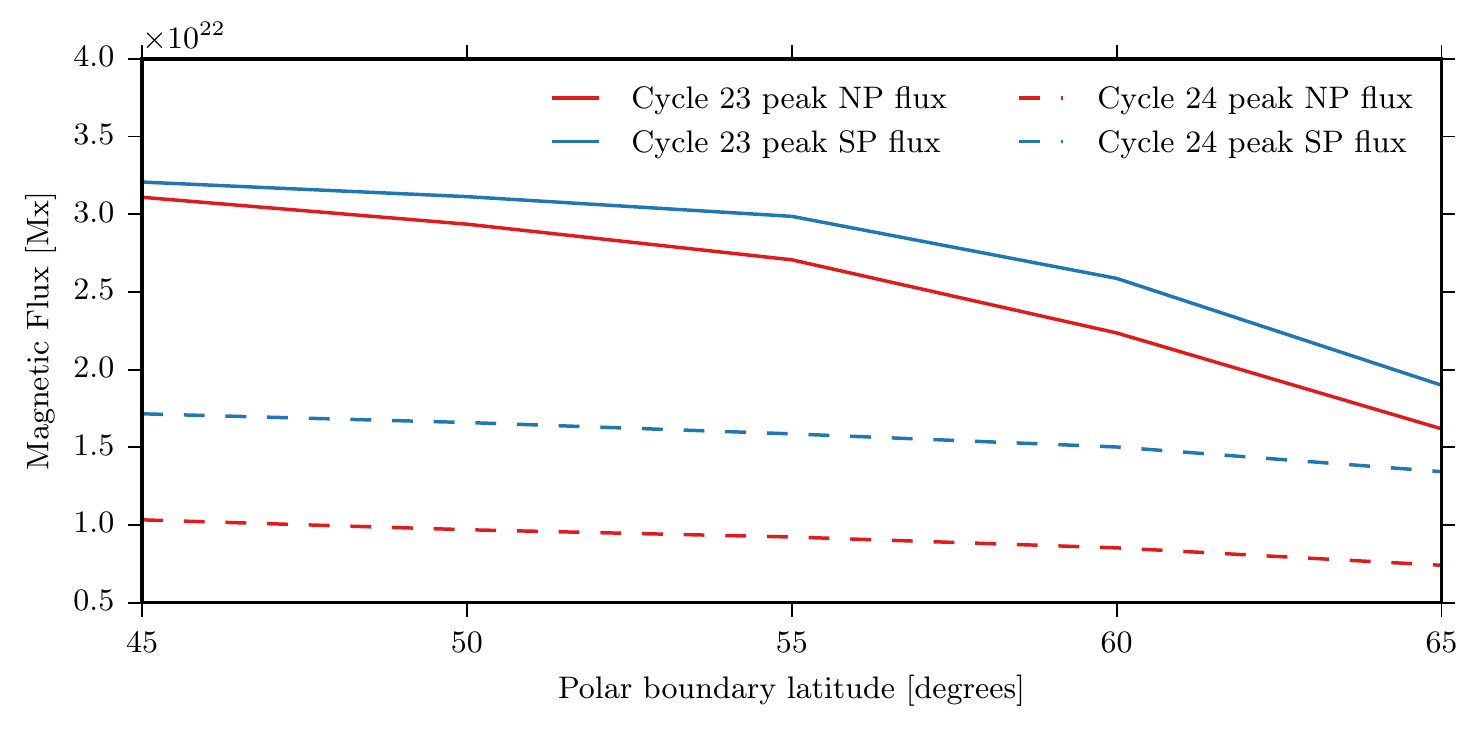}
\caption{Peak magnitude of coronal hole enclosed polar magnetic flux for each pole, as a function of latitude for the polar boundaries defined from $\pm$\{45, 50, 55, 60, 65\} degrees. The northern and southern polar values are mapped in red and blue, respectively. Solar Cycle 23 data is marked with solid lines, while solar Cycle 24 data is mapped in a dashed fashion. As the polar boundary is pushed further poleward in each direction, this enclosed magnetic flux reduces in magnitude.}
\label{fig:pflux-comp-varlat}\end{center}
\end{figure*}

To quantify the effects of this boundary shift, a profile of open magnetic flux was created for each polar region as defined by the ranges of boundary latitudes previously defined. From each of these profiles, the peak value of open magnetic flux was computed for the range of solar Cycles 23 and 24 separately. Figure~\ref{fig:pflux-comp-varlat} plots these peak values of polar open magnetic flux as a function of the choice of polar latitude boundary. The northern and southern polar regions are plotted in red and blue, respectively. Solar Cycle 23 peaks are plotted in solid curves, with Cycle 24 shown by dashed curves. As the polar boundary is pushed further poleward, peak polar open fluxes decline as more enclosed flux is re-classified as low-latitude. For both the northern and southern hemispheres of solar Cycle 23, a more significant drop in peak polar open magnetic flux occurs as the polar boundary pushes beyond $\pm$ 55 degrees latitude towards the poles. This same trend does not appear for the portion of Cycle 24 data gathered. For the coronal hole distributions of Cycle 23, moving poleward from $\pm$ 55 degrees latitude begins a significant drop in peak open magnetic flux.

These analyses suggest that a polar boundary of $\pm$ 55 degrees latitude provides a balanced choice. It allows for capture of polar open flux during more standard periods of coronal hole activity, while also providing a buffer to capture extensions into slightly lower latitudes without being integrated to lower latitude profiles.

%
\begin{acks}
This work was supported by the NASA Living with a Star (LWS) program. This work utilizes datasets available from the SDO, STEREO, and SOHO missions. Particular thanks to Dana Longcope and Richard Canfield for discussion of this project. We also thank the referee for constructive questions and comments, which have clarified and improved this work.
\end{acks}

\begin{disc}
The authors declare that they have no conflicts of interest.
\end{disc}

%
%
\bibliographystyle{spr-mp-sola}
\bibliography{gachd}

\end{article}
\end{document}